\documentclass[%
 ,secnumarabic%
,amssymb, amsmath,nobibnotes, aps, pre]{revtex4}
\usepackage[dvipdfmx]{graphicx}
\usepackage{latexsym,mathrsfs}
\usepackage{dcolumn}
\usepackage{bm}

\def \ve#1{\mbox{\boldmath $#1$}}

\begin{document}
\title{Neo-logistic model for the growth of bacteria}

\author{Tohru Tashiro\footnote{tashiro@cosmos.phys.ocha.ac.jp}}
  \affiliation{Department of Physics, Ochanomizu University,
	       2-1-1 Ohtuka, Bunkyo, Tokyo 112-8610, Japan}

\author{Fujiko Yoshimura}
  \affiliation{Department of Information Sciences, Ochanomizu University,
	       2-1-1 Ohtuka, Bunkyo, Tokyo 112-8610, Japan}

\date{\today}

\begin{abstract}
We propose a neo-logistic model that can describe bacterial growth data precisely. 
This model is not derived by modifying the logistic model formally, but by incorporating the synthesis of inducible enzymes into the logistic model indirectly. Therefore, the meaning of the parameters of the neo-logistic model becomes physically clear. 
The neo-logistic model can approximate bacterial growth better than models previously presented, and predict the order of the saturated number of bacteria in the stationary phase from initial data including, and just after the end of, the lag phase much more accurately.
\end{abstract}

\maketitle

\section{Introduction}
\label{sec:intro}
Regardless of bacteria type, if the bacteria are enclosed in a liquid medium with a single type of nutrient (batch culture \cite{Heritage96}), then they exhibit the following universal growth: In the initial period, called the {\it lag phase}, the bacteria are adjusting to their new environment, and do not divide. After the lag phase, the bacteria have adjusted and begin to divide. This period is called the {\it exponential phase}, as the growth increases exponentially. The next phase is called the {\it stationary phase}, where cell division stops and the number of bacteria becomes saturated. The last phase is the {\it death phase}, where the number of bacteria declines \cite{Heritage96}.
Interestingly, if the death phase is neglected, the temporal change of the logarithmic number of bacteria can be simply characterized by an S-shaped curve \cite{Hauschild92}, even though the biochemical reactions utilizing the nutrient as a substrate for energy production are complex, and depend on the type of nutrient used.

There should be dynamical systems that can be used to describe this S curve behavior.
In predictive microbiology, the descriptions of bacterial growth by mathematical models are used to describe growth of the microorganisms, and to investigate how growth is influenced by the surrounding environment. Several models to describe bacterial growth have been derived by modifying existing models, e.g. the logistic model (LM) \cite{Verhulst38,Verhulst45,Verhulst47} and the Gompertz model (GM) \cite{Gompertz25}, and these models have also been used to describe the growth in organisms other than bacteria. 

LM can be derived from Malthusian law by adding an upper limit to the growth, and it is then represented by the following differential equation:
\begin{equation}
\frac{{\rm d}N^{\rm L}(t)}{{\rm d}t} = a\left\{N^{\rm L}_\infty-{N^{\rm L}(t)}\right\}N^{\rm L}(t) \ ,
\label{eq:LM}
\end{equation}
where $N^{\rm L}(t)$ is the number of an organism at time $t$. 
When $N^{\rm L}(t)$ becomes closer to $N^{\rm L}_\infty$ the right hand side of Eq.~(\ref{eq:LM}) converges to 0, which means that $N^{\rm L}_\infty$ is the saturated number.
The solution of Eq.~(\ref{eq:LM}) is
\begin{equation}
N^{\rm L}(t) = \frac{N^{\rm L}_\infty}{1+\left(\frac{N^{\rm L}_\infty}{N^{\rm L}(0)}-1\right){\rm e}^{-a N^{\rm L}_\infty t}} \ .
\end{equation}
GM was originally developed as a model for population decline reflecting the statistics that indicate a death rate increase with age. Then, GM began to be used as a model for population growth of an organism, by changing the sign of the coefficient of time \cite{Wright26,Davidson28,Weymouth31}.
GM can be represented by
\begin{equation}
N^{\rm G}(t) = C\exp\left[-\exp\left\{-D(t-M)\right\}\right] \ ,
\end{equation}
where $N^{\rm G}(t)$ is the number of an organism at time $t$.

In adopting these models for describing bacterial growth, we are faced with an essential problem that they cannot describe the temporal change of the {\it logarithmic} number of bacteria, which follows the S-shaped curve. Therefore, these models cannot be utilized to describe real data of bacteria growth without modification.

One can easily show why neither of these models exhibits the S-shaped curve on the logarithmic scale. In order for a monotonously increasing function in time that converges as $t\rightarrow\infty$ to assume an S-shaped curve in the logarithmic scale, it must have an inflection point at some finite time. However, the second derivatives of these models can be calculated as
\begin{align}
\frac{{\rm d}^2}{{\rm d}t^2}\ln N^{\rm L}(t) &= -a{N^{\rm L}(t)}^2\left(\frac{N^{\rm L}_\infty}{N^{\rm L}(0)}-1\right){\rm e}^{-a N^{\rm L}_\infty t}  \ , \\
\frac{{\rm d}^2}{{\rm d}t^2}\ln N^{\rm G}(t) &= -D^2\exp\left\{-D(t-M)\right\}   \ ,
\end{align}
and they are always negative. 

One method to modify these models so that they may be used to analyze real data of bacterial growth, is to take the exponential of the models:
\begin{equation}
\frac{B^{\rm mL}(t)}{B^{\rm mL}(0)} = \exp\left[N^{\rm L}(t)-N^{\rm L}(0)\right] \ \Longrightarrow \ {B^{\rm mL}(t)} = {B^{\rm mL}(0)}\exp\left[-\frac{\beta}{1+\delta}+\frac{\beta}{1+\delta{\rm e}^{-\gamma t}}\right] \ ,
\label{eq:MLM}
\end{equation}
\begin{equation}
\frac{B^{\rm mG}(t)}{B^{\rm mG}(0)} = \exp\left[N^{\rm G}(t)-N^{\rm G}(0)\right] \ \Longrightarrow \ {B^{\rm mG}(t)} = {B^{\rm mG}(0)}\exp\left[-C\exp\left[-{\rm e}^{DM}\right]+C\exp\left[-\exp\left\{-D(t-M)\right\}\right]\right] \ ,
\label{eq:MGM}
\end{equation}
where $B^{\rm mL}(t)$ and $B^{\rm mG}(t)$ represent the number of bacteria in the modified logistic model (MLM) and the modified Gompertz model (MGM), respectively. The parameters in MLM relate to those in LM as $\beta=N^{\rm L}_\infty$, $\gamma = a{N^{\rm L}_\infty}$, and $\delta={N^{\rm L}_\infty}/{N^{\rm L}(0)}-1$.
Both MLM and MGM have been utilized extensively for predicting the growth of bacteria \cite{Gibson87,Kacena99,Shi03,Ponce08,Szczawinsk14,Zhou15,Charteris01,Kayombo03,Lavelli06,Huang10,Avila-Sosa10,Tomac13,Dalcanton13,Alonso-Hernando13,Kalschne14,Sagdic14,Dotto15,Ozdemir17}.

However, while these modified models can approximate the behavior of bacterial growth formally, 
there is no correlation between these models and the actual mechanisms of bacterial growth. 
Therefore, in this paper we construct a simple model describing bacterial growth precisely, by incorporating the realistic behavior of bacteria. 

There are a large number of biochemical processes that occur in the growth of bacteria, and a theory that describes the transitions between the lag, exponential, stationary, and death phases of bacteria growth precisely, by considering the autocatalytic processes, which includes substrates, active chemical components, non-growth-facilitating components, and the complexes of these  \cite{Himeno17}. 
However, here we aim to construct a much simpler model that can realistically describe bacterial growth through the lag, exponential, and stationary phases in the batch culture with fewer variables. The approach we employ takes into account only the relationship between bacteria and substrates. Accordingly, we chose to modify LM for this purpose. The meanings of the parameters in this new model are physically clear. Moreover, we have used this model to predict the saturated number of bacteria in the stationary phase from 
initial data including, and just after the end of, the lag phase.

This paper is organized as follows. In Sec.~\ref{sec:logi}, LM is reconsidered and is derived in an original way. 
By doing this, we can understand the correlation between LM and real processes, however, the processes behind LM do not include those involving inducible enzymes. Therefore, in Sec.~\ref{sec:neologi}, we construct a new and simpler modified LM, the neo-logistic model (NLM), taking into account the synthesis of inducible enzymes. In Sec.~\ref{sec:convegence}, we derive the convergence value of the total number of the bacteria in the long-time limit using NLM. Then, we fit real bacterial growth data using NLM, and compare the agreement of NLM with the agreement of previous models in Sec.~\ref{sec:fitting}. Finally, in Sec.~\ref{sec:prediction}, we compare the accuracy of NLM and previous models in predicting the order of the saturated number of bacteria from initial data including, and just after the end of, the lag phase.

\section{Reconsideration of the logistic model}
\label{sec:logi}

In this section we reconsider LM to find a correlation with actual mechanisms in bacterial growth. 
The general interpretation of LM is that it models the population growth of an organism in a similar way to Malthusian law, but with an additional term which reflects the slowing in the growth of an organism due to its limited living space \cite{Braun13}. 
From this interpretation, however, we cannot develop the model with regard to finding a correlation to real mechanisms in bacterial growth. 
Therefore, we derive LM in an original way, paying attention only to the relationship between bacteria and substrates. 

Consider a closed three-dimensional space where a single species of bacteria and substrates exist, i.e., the batch culture. 
The space is a cube whose sides are of length $L$ for simplicity. 
In order to coarse-grain the behavior of bacteria, we shall divide the space into $n^3$ cubes with length $\Delta l (=L/n)$, and discretize the time with interval $\Delta t$. We then suppose that a bacterium can be represented as $\eta$ unit cubes, and the average substrates required for one cell division can be represented as a single unit cube. That is, the volumes of the bacterium and the average substrates required for cell division are $\eta\Delta l^3$ and $\Delta l^3$, respectively.

As bacteria swim around in the culture they absorb the substrates required for cell division. However, we idealize this circumstance, and assume that the average amount of substrates required for cell division can be localized in a unit cube. Although this idealization seems oversimplified, it is required for our derivation of LM.

For this system we shall employ the following rules which are working hypotheses for deriving LM: (i) The bacterium move to the next unit cube randomly in each discretized time step $\Delta t$ without overlapping. (ii) When the bacterium comes in contact with a substrate cube, it absorbs that cube (the substrate cube disappears) and divides (the bacterium becomes two separate bacterium occupying $2\eta$ unit cubes). In addition we set the number of bacteria and substrate cubes at the $i$th time step as $B_i$ and $S_i$, respectively.

We calculate the probability of a bacterium and a substrate cube making contact by adopting a mean-field-like approach. 
We assume that they are distributed equally throughout the space, and both the probability of a bacterium making contact with a substrate cube, and the probability of a substrate cube making contact with a bacterium have constant values throughout the space for any time interval $\Delta t$. 
It follows that the probability of a bacterium meeting a substrate cube is proportional to the number of substrate cubes, and the probability of a substrate cube making contact with a bacterium is proportional to the number of bacteria. When considering the number of cubes a bacterium occupies, the probability of the bacterium making contact with a substrate cube, and a substrate cube making contact with a bacterium can be represented by $\eta{B_i}/{n^3}$ and $\eta{S_i}/{n^3}$, respectively. 

According to rule (ii), the number of bacteria that divide is $\eta{S_i}B_i/{n^3}$, and the number of substrate cubes that are absorbed is $\eta{B_i}S_i/{n^3}$. The change in the number of the bacteria and substrate cubes can be represented by the following recursion formula:
\begin{align}
B_{i+1} &= \left(1-\eta\frac{S_i}{n^3}\right)B_i + {2}\times\eta\frac{S_i}{n^3}B_i = B_i + \eta\frac{S_i}{n^3}B_i  \ , \\
S_{i+1} &= S_i - \eta\frac{B_i}{n^3}S_i \ .
\end{align}

We shall define the total number of bacteria and substrate cubes at $t$ as $B^{\rm L}(t) = B^{\rm L}(i\cdot\Delta t) \equiv B_i$ and $S^{\rm L}(t) = S^{\rm L}(i\cdot\Delta t)\equiv S_i$, respectively. 
We move $B_i$ and $S_i$ to the other sides each equation, divide both equations by $\Delta t$, and take the limits as $n\rightarrow\infty$ and $\Delta t\rightarrow0$ with $n^3\Delta t$ fixed. 
By setting this fixed value as
\begin{equation}
\frac{a}{\eta} = \lim_{\stackrel{\scriptstyle n\rightarrow \infty}{\Delta t\rightarrow0}}\frac{1}{n^3\Delta t} \ , 
\end{equation}
we can obtain the following differential equations:
\begin{align}
\frac{{\rm d}B^{\rm L}(t)}{{\rm d}t} &= aS^{\rm L}(t)B^{\rm L}(t) \label{eq:LM_bac}  \ , \\
\frac{{\rm d}S^{\rm L}(t)}{{\rm d}t} &=-aB^{\rm L}(t)S^{\rm L}(t) \ . \label{eq:LM_bac2}
\end{align}

We find that the total amount of bacteria and substrate cubes is conserved from these equations. By setting the conserved quantity as $B^{\rm L}_\infty\equiv B^{\rm L}(t)+S^{\rm L}(t)$ and substituting this into Eq.~(\ref{eq:LM_bac}), we obtain the following equation:
 \begin{equation}
\frac{{\rm d}B^{\rm L}(t)}{{\rm d}t} = a\left\{B^{\rm L}_\infty - B^{\rm L}(t)\right\}B^{\rm L}(t) \ ,
\end{equation}
which represents LM.

\section{Derivation of the neo-logistic model}
\label{sec:neologi}

In the previous section, LM was derived from the elementary process of bacteria absorbing a substrate cube resulting in the cell division. 
As mentioned in Sec.~\ref{sec:intro}, LM does not describe the S-shaped curve of bacterial growth on the logarithmic scale. 
That is, the lag phase of the LM curve disappears on the logarithmic scale. Therefore, we shall construct a new modified LM, NLM, by adding an additional rule.

Bacteria obtain the energy for biological activities including cell division from biochemical reactions that convert substrate with enzymes. 
The role of enzymes in these reactions is quite important. 
The velocity of biochemical reactions without enzymes increases $10^7\mbox{--}10^{20}$ fold with the addition of enzymes \cite{Voet90}. 
Enzymes have a specificity for substrates and enhance the rate of essential biochemical reactions with those substrates. 

In the lag phase, the bacteria produce an inducible enzyme specialized for the substrate, which greatly increases the rate of biochemical reactions by which cells obtain the energy for cell division \cite{Heritage96}.
If bacteria that are producing inducible enzyme are then introduced into another identical culture, their lag time is decreased. 
In addition, a second lag phase can occur in a culture that has two types of nutrients (substrates), e.g. glucose and lactose. 
Bacterial growth stagnates when bacteria exhaust the preferred nutrient source (glucose), and then the bacteria begin to produce an enzyme specialized for the non-preferred nutrient (lactose) and enter a second lag phase called the  {\it diauxic} lag phase \cite{Monod49,Stanier51}. 
This lag phase is also understood as a period when an enzyme specialized for the non-preferred is produced \cite{Stanier51}.

We shall incorporate these processes in the rule (ii). 
In this way, we avoid introducing a new variable representing the inducible enzyme, and deal only with the bacteria and substrates.

The new rule (ii) is as follows: In a growth culture with one type of substrate, while a bacterium absorbs a substrate cube $m$ times, the bacterium produce an inducible enzyme particular to the substrate. When possessing the inducible enzyme, the cell can immediately devide after it absorbs a substrate cube.
Then, when this cell divides, the daughter cells also possess this enzyme, i.e., they can double just after absorbing a substrate cube.

We shall call the number of substrate cubes the bacterium has absorbed from time $t=0$ to the first cell division the cell's {\it rank}. A bacterium begins with a rank of 0, and its rank increases by 1 after absorbing each additional substrate cube. Therefore, as the bacteria absorb substrate cubes, the number of bacteria of rank 0 declines. On the other hand, the number of bacteria with rank $k$, where $1\le k\le m-1$, does not continuously decrease because bacteria with rank $k$ are produced from bacteria with rank $k-1$. 
After a bacterium with rank $m$ absorbs an additional substrate cube, it immediately divides, and the daughter cells retain a rank of $m$. Let us set the number of bacteria with a rank of $k$ at the $i$th time step as $B_i^k$.
Then the recursion formula can be written as
\begin{align}
B_{i+1}^m &= \left(1-\eta\frac{S_i}{n^3}\right)B_i^m + 2\times\eta\frac{S_i}{n^3}B_i^m+\eta\frac{S_i}{n^3}B_i^{m-1} 
 = B_i^m + \eta\frac{S_i}{n^3}B_i^m+ \eta\frac{S_i}{n^3}B_i^{m-1}  \ ,\nonumber \\
B_{i+1}^{m-1}  &=  \left(1-\eta\frac{S_i}{n^3}\right)B_i^{m-1}+ \eta\frac{S_i}{n^3}B_i^{m-2} \ ,\nonumber \\
&\ \ \vdots \label{re:neologi} \\
B_{i+1}^{0}  &=  \left(1-\eta\frac{S_i}{n^3}\right)B_i^{0} \ ,\nonumber \\
S_{i+1} &= S_i - \eta\sum_{k=0}^{m}\frac{B_i^k}{n^3}S_i  \ .\nonumber
\end{align}
Taking the limits as $n\rightarrow\infty$ and $\Delta t\rightarrow0$, we can obtain the following differential equations:
\begin{align}
\frac{{\rm d}B^{m}(t)}{{\rm d}t} &= aS(t)B^{m}(t) + aS(t)B^{m-1}(t) \nonumber \\
\frac{{\rm d}B^{m-1}(t)}{{\rm d}t} &= -aS(t)B^{m-1}(t) + aS(t)B^{m-2}(t) \nonumber \\
&\ \ \vdots \label{eq:neologi} \\
\frac{{\rm d}B^{0}(t)}{{\rm d}t} &= -aS(t)B^{0}(t)  \nonumber \\
\frac{{\rm d}S(t)}{{\rm d}t} &= -a\sum_{k=0}^{m}B^{k}(t)S(t) \, \nonumber
\end{align}
with the initial conditions
\begin{equation}
B^0(0) \equiv B_0, B^k(0) = 0 \ (k=1, 2, \cdots, m) \ {\rm and} \ S(0) \equiv S_0 \ ,
\end{equation}
where $B^k(t) = B^k(i\cdot\Delta t) \equiv B_i^k$ and $S(t) = S(i\cdot\Delta t) \equiv S_i$. 
We shall call these equations the {\it neo-logistic model} (NLM). 
In addition, we will now denote the total number of bacteria at time $t$ as
\begin{equation}
B(t) = \sum_{k=0}^{m}B^k(t) \ .
\end{equation}

When $m=0$, this model changes into
\begin{equation}
\frac{{\rm d}B(t)}{{\rm d}t} = aS(t)B(t)  \ , \ \frac{{\rm d}S(t)}{{\rm d}t} = -aB(t)S(t) \ ,
\end{equation}
which coincides with Eqs.~(\ref{eq:LM_bac}) and (\ref{eq:LM_bac2}), and so NLM includes LM.

When $m > 0$, the derivative of $B(t)$ with respect to time can be derived by summing up $\dot{B}^k(t)$ of Eqs.~(\ref{eq:neologi}) for all $k$, and can be expresses as $\dot{B} (t) = aS(t)B^m(t)$. For a finite $t$, $B^m(t)>0$, and $S(t)>0$ resulting in $\dot{B}(t)>0$. 
However, $\dot{B}(0)=0$ due to the initial conditions. In addition, $\dot{B}(t) \rightarrow 0$ for $t \rightarrow \infty$ since the bacteria consume all of the substrates in this limit. Therefore, $\frac{\rm d}{{\rm d}t}\ln B(t)=\dot{B}(t)/B(t)$ has a maximum value at a a finite time $t$, which implies an inflection point exists, contrary to LM and GM. Therefore, the total number of the bacteria of NLM exhibit the S-shaped curve in a semi-logarithmic plot.

\begin{figure}[h]
  \begin{center}
      \includegraphics[scale=0.9]{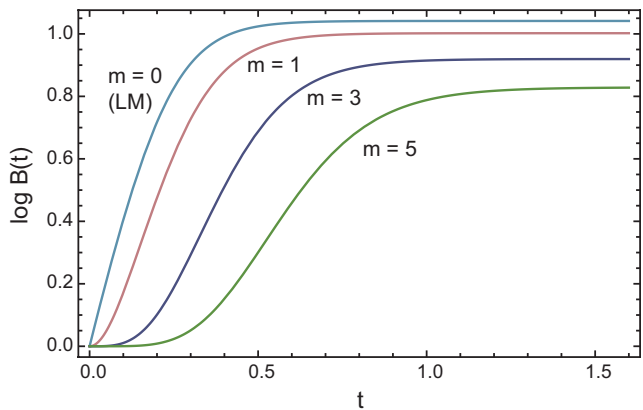}
    \caption{\label{fig:result}(color online) The total number of bacteria in NLM as a function of time for $m = 0$, 1, 3, and 5 with $a = 1$, $B_0 = 1$, and $S_0 = 10$. The case with $m = 0$ corresponds to LM.}
  \end{center}
\end{figure}

The time evolution of the total number of bacteria in NLM for $m = 0$, 1, 3 and 5 with $a = 1$, $B_0 = 1$, and $S_0 = 10$ are shown in Fig.~\ref{fig:result}. The case with $m = 0$ corresponds to LM. We can see that NLM with $m > 0$ describes the lag phase in the semi-logarithmic plot, which LM cannot do, and the interval of this phase increases with increasing $m$. 

As will be clear in Sec.~\ref{sec:fitting}, in real experimental data, the number of bacteria are indicated in terms of density, or the number of bacteria per unit volume. Thus we will rewrite NLM by using the number of bacteria with a rank of $k$, and the number of substrate cubes per unit volume as follows.
\begin{equation}
b^k(t) \equiv \frac{B^k(t)}{L^3} \ (k=1,2,\cdots,m) \ \mbox{and} \ s(t) \equiv \frac{S(t)}{L^3} \ .
\end{equation}

By dividing both sides of Eqs.~(\ref{eq:neologi}) by $L^3$ we can obtain
\begin{align}
\frac{{\rm d}b^{m}(t)}{{\rm d}t} &= \alpha s(t)b^{m}(t) + \alpha s(t)b^{m-1}(t) \nonumber \\
\frac{{\rm d}b^{m-1}(t)}{{\rm d}t} &= -\alpha s(t)b^{m-1}(t) + \alpha s(t)b^{m-2}(t) \nonumber \\
&\ \ \vdots \label{eq:neologi3} \\
\frac{{\rm d}b^{0}(t)}{{\rm d}t} &= -\alpha s(t)b^{0}(t)  \nonumber \\
\frac{{\rm d}s(t)}{{\rm d}t} &= -\alpha\sum_{k=0}^{m}b^{k}(t)s(t) \ , \nonumber
\end{align}
with the initial conditions
\begin{equation}
b^0(0) \equiv b_0, b^k(0) = 0 \ (k=1, 2, \cdots, m) \ {\rm and} \ s(0) \equiv s_0 \ ,
\end{equation}
where the parameter $\alpha$ is defined as 
\begin{equation}
\alpha \equiv L^3a = \lim_{\stackrel{\scriptstyle n\rightarrow \infty}{\Delta t\rightarrow0}}\frac{\eta L^3}{n^3\Delta t} \ . 
\label{eq:def_alpha}
\end{equation}

Finally, we shall investigate the physical meaning of $\alpha$.
By using $L=n\Delta l$, the equation~(\ref{eq:def_alpha}) can be written as
\begin{equation}
\alpha = \lim_{\stackrel{\scriptstyle \Delta l\rightarrow 0}{\Delta t\rightarrow0}}\frac{\eta \Delta l^3}{\Delta t} \ . 
\label{eq:def_alpha2}
\end{equation}

In the above expression, $\eta\Delta l^3$ is the volume of a single bacterium. Moreover,  $\Delta t$ can be exchanged with ``(length of a substrate cube)/(velocity of bacteria)" because the bacteria move $\Delta l$ which is the length of a substrate cube within this time interval. Therefore, we can interpret the meaning of $\alpha$ as
\begin{equation}
\alpha \sim \frac{
(\mbox{volume of bacteria})\times(\mbox{velocity of bacteria})
}{
(\mbox{length of a substrate cube})
} \ .
\label{eq:interpretation_of_alpha}
\end{equation}

\section{Convergence value of the model}
\label{sec:convegence}

Now we shall investigate the convergence value of $B(t)$ in the long-time limit. The convergence value of NLM with $m = 0$, i.e. LM, can be derived as
\begin{equation}
B^{\rm L}(\infty) = B^{\rm L}(0) + S^{\rm L}(0) \ ,
\label{eq:conv_logi}
\end{equation}
where we have used $B^{\rm L}(t)+S^{\rm L}(t)=\mbox{const.}$ and $S^{\rm L}\rightarrow0$ at $t\rightarrow\infty$.
In contrast, we cannot obtain the convergence value for NLM with $m>1$ easily, because the number of bacteria and substrates are not conserved. 

We shall introduce a {\it new time} $\tau$ defined as
\begin{equation}
\tau(t) = \int_{0}^{t}aS(t'){\rm d}t' \ .
\end{equation}
Here, $\tau(0)=0$.  Because $aS(t)\ge0$, $\tau$ is a monotonically increasing function of $t$.

By using $\tau$, NLM as represented by Eqs.~(\ref{eq:neologi}) becomes
\begin{align}
\frac{{\rm d}\tilde{B}^{m}(\tau)}{{\rm d}\tau} &= \tilde{B}^{m}(\tau) + \tilde{B}^{m-1}(\tau) \nonumber \\
\frac{{\rm d}\tilde{B}^{m-1}(\tau)}{{\rm d}\tau} &= -\tilde{B}^{m-1}(\tau) + \tilde{B}^{m-2}(\tau) \nonumber \\
&\ \ \vdots \label{eq:neologi2}\\
\frac{{\rm d}\tilde{B}^{1}(\tau)}{{\rm d}\tau} &= -\tilde{B}^{1}(\tau) + \tilde{B}^{0}(\tau) \nonumber \\
\frac{{\rm d}\tilde{B}^{0}(\tau)}{{\rm d}\tau} &= -\tilde{B}^{0}(\tau)  \nonumber
\end{align}
and
\begin{align}
\frac{{\rm d}\tilde{S}(\tau)}{{\rm d}\tau} &= -\sum_{k=0}^{m}\tilde{B}^{k}(\tau) \ ,
\label{eq:S}
\end{align}
where $\tilde{B}^k(\tau)=\tilde{B}^k[\tau(t)]\equiv{B}^k(t) \ (k=0,1,\cdots, m) $ and $\tilde{S}(\tau)=\tilde{S}[\tau(t)]\equiv{S}(t)$.

The equations~(\ref{eq:neologi2}) are homogeneous equations for $\tilde{B}^k(\tau)$ with constant coefficients, so we can
solve them analytically. First, we shall rewrite them as
\begin{equation}
\frac{{\rm d}}{{\rm d}\tau}\tilde{\ve{B}}_m(\tau) = \ve{A}_m\tilde{\ve{B}}_m(\tau)  \ ,
\end{equation}
where
\begin{equation}
\tilde{\ve{B}}_m(\tau) = (\tilde{B}^{m}(\tau),\cdots,\tilde{B}^1(\tau),\tilde{B}^0(\tau))^{\rm T}
\end{equation}
and
\begin{equation}
\ve{A}_m = \left(
\begin{array}{cccccc}
1 & 1 & 0 & 0 & \cdots & 0 \\
0 &-1 & 1 & 0 & \cdots & 0 \\
0 & 0 &-1 & 1 & \cdots & 0 \\
\vdots&\vdots&\vdots&\vdots&\ddots&\vdots \\
0 & 0 & 0 & 0 & \cdots & -1 
\end{array}
\right) \ .
\end{equation}

The solution of this equation with the initial conditions $\tilde{B}^k(0)={B}^k(0)=\delta_{k,0}B_0$ where $\delta_{i,j}$ is the Kronecker delta is
\begin{equation}
\tilde{\ve{B}}_m(\tau) = \exp\left[\ve{A}_m\tau\right]\tilde{\ve{B}}_m(0) = \left(\frac{{\rm e}^{\tau}}{2^{m}}\left\{1-\frac{\Gamma(m,2\tau)}{m!}\right\},\frac{\tau^{m-1}{\rm e}^{-\tau}}{(m-1)!}\cdots,\tau{\rm e}^{-\tau},{\rm e}^{-\tau}\right)^{\rm T}B_0
\end{equation}
where $\Gamma(m,\tau)$ is the incomplete Gamma function.
Then, the sum of $\tilde{B}^k(\tau)$ becomes
\begin{align}
\sum_{k=0}^{m}\tilde{B}^k(\tau) &= \left[\sum_{k=0}^{m}\frac{\tau^{k}{\rm e}^{-\tau}}{k!}+\frac{{\rm e}^{\tau}}{2^{m}}\left\{1-\frac{\Gamma(m,2\tau)}{m!}\right\}\right]B_0 \nonumber \\
 &= \left[\frac{\Gamma(m,\tau)}{(m-1)!}+\frac{{\rm e}^{\tau}}{2^{m}}\left\{1-\frac{\Gamma(m,2\tau)}{(m-1)!}\right\}\right]B_0 \nonumber \\
&\equiv F_m(\tau)B_0 \ .
\end{align}
The initial value of $F_m$ defined above is
\begin{equation}
F_m(0) = \frac{\Gamma(m)}{(m-1)!}+\frac{1}{2^{m}}\left\{1-\frac{\Gamma(m)}{m!}\right\} = \frac{(m-1)!}{(m-1)!}+\frac{1}{2^{m}}\left\{1-\frac{(m-1)!}{(m-1)!}\right\} = 1 \ ,
\end{equation}
here we have used $\Gamma(m,0)=\Gamma(m)$ where $\Gamma(m)$ is the Gamma function.

 We can derive $\tilde{S}(\tau)$ by integrating Eq.~(\ref{eq:S}) with the initial condition $\tilde{S}(0)={S}(0)=S_0$ as
\begin{equation}
\tilde{S}(\tau) = S_0 - B_0\int_{0}^{\tau}F_m(\tau'){\rm d}\tau' \ .
\label{eq:rel_S0_B0}
\end{equation}
Let us define $G_m$ as 
\begin{equation}
G_m(\tau) = \int_{0}^{\tau}F_m(\tau'){\rm d}\tau'  = m-1 + \frac{(\tau+1)\Gamma(m,\tau)-\Gamma(m+1,\tau)}{(m-1)!} + \frac{{\rm e}^{\tau}}{2^{m}}\left\{1-\frac{\Gamma(m,2\tau)}{(m-1)!}\right\} \ .
\label{eq:Gm}
\end{equation}
Here, $G_m(0)=0$. Because $G_m'=F_m>0$, the inverse function for $G_m$ exists.

Consider the equations at $t\rightarrow\infty$. We shall define
\begin{equation}
\tau_\infty = \tau(\infty) = \int_{0}^{\infty}aS(t){\rm d}t \ .
\end{equation}
In this limit, the bacteria consume all of the substrate cubes, i.e., $\tilde{S}(\tau_\infty) = S(\infty) = 0$, so the number of bacteria assumes a convergence value, which means the system reaches a stationary phase. Therefore, this convergence value which is defined as
 \begin{equation}
B_\infty\equiv\sum_{k=0}^{m}B^k(\infty) = \sum_{k=0}^{m}\tilde{B}^k(\tau_\infty)
\end{equation}
satisfies the following equation
\begin{equation}
  B_\infty = F_m(\tau_\infty)B_0  \ , \label{eq:Binf}
\end{equation}
and equation~(\ref{eq:rel_S0_B0}) becomes
\begin{equation}
S_0 = G_m(\tau_\infty)B_0  \ . \label{eq:S0}
\end{equation}

From Eq.~(\ref{eq:S0}), using the inverse function for $G_m$, $\tau_\infty$ can be represented as $\tau_\infty = G_m^{-1}(S_0/B_0)$. Substituting this relation into Eq.~(\ref{eq:Binf}), we can express the convergence value as
\begin{equation}
B_\infty = F_m\left[G_m^{-1}\left(\xi\right)\right]B_0
\label{eq:conv}
\end{equation}
where we define $\xi\equiv S_0/B_0$.

With regard to  $m=1$, because $F_1(\tau_\infty)=\cosh\tau_\infty$, and $G_1^{-1}(\xi)={\rm arcsinh}\xi$, we find
\begin{equation}
\frac{B_\infty}{B_0} = \sqrt{1+\xi^2} \ .
\end{equation}
On the other hand, for a general value of $m$, it is impossible to obtain such a brief expression of the convergence value as a function of $\xi$.
Thus, we shall obtain asymptotic expressions of the convergence value for $\xi\ll1$ or $\xi\gg1$. 

For small values of $\xi$, we obtain
\begin{align}
F_m(\tau)&= 1 + \frac{1}{(m+1)!}\tau^{m+1} + (\tau^{m+2}) \ , \\
G_m^{-1}(\xi) &= \xi + O(\xi^2) \ ,
\end{align}
where we have used the expansion of $\Gamma(m,\tau)$:
\begin{equation}
\Gamma(m,\tau) = m! - \frac{\tau^{m+1}}{m+1} + \frac{\tau^{m+2}}{m+2} +O(\tau^{m+3}) \ .
\end{equation}
With regard to small $\xi$, therefore, we can obtain
\begin{equation}
\frac{B_\infty}{B_0} \simeq 1 + \frac{1}{(m+1)!}\xi^{m+1} \ .
\label{eq:Binf_small}
\end{equation}

Next, let us consider the case where  $\xi\gg1$. When we fit real data for the change in the number of bacteria in time using NLM, the value of $S_0/B_0$ is frequently quite large, which will be made clear in Sec.~\ref{sec:fitting}. 
From Eqs.~(\ref{eq:Gm}) and (\ref{eq:S0}), 
\begin{equation}
\xi = G_m(\tau_\infty) = m-1 + \frac{(\tau_\infty+1)\Gamma(m,\tau_\infty)-\Gamma(m+1,\tau_\infty)}{(m-1)!} + \frac{{\rm e}^{\tau_\infty}}{2^{m}}\left\{1-\frac{\Gamma(m,2\tau_\infty)}{(m-1)!}\right\} \ .
\label{eq:Gm2}
\end{equation}
In order that this relation holds for $\xi\gg1$, $\tau_\infty$ also must be large because $G_m$ is a monotonically increasing function. 
For large $\tau_\infty$, the term ${\rm e}^{\tau_\infty}/2^{m}$ remains on R.H.S of the relation, and the other terms cancel out, so that
 \begin{equation}
\xi \simeq \frac{{\rm e}^{\tau_\infty}}{2^{m}} \ ,
\label{eq:Gm3}
\end{equation}
and then
\begin{equation}
\tau_\infty\simeq \ln(2^{m}\xi) \ ,
\label{eq:tau}
\end{equation}
where we have used the asymptotic formulation of $\Gamma(m,\tau)$ with large $\tau$ \cite{Abramowitz72}:
\begin{equation}
\Gamma(m,\tau) \simeq \tau^{m-1}{\rm e}^{-\tau} \ .
\label{eq:Gamma_asy}
\end{equation}
Hence, we can obtain the asymptotic expression of $B_\infty/B_0$ for $\xi\gg1$ as
\begin{equation}
\frac{B_\infty}{B_0} \simeq \left\{1-\frac{\Gamma\left[m,2\ln(2^{m}\xi)\right]}{(m-1)!}\right\}\xi \ .
\label{eq:Binf_large}
\end{equation}

By letting $m$ go to 0, we can see that both asymptotic expressions Eqs.~(\ref{eq:Binf_small}) and (\ref{eq:Binf_large}) coincide with the logistic model Eq.~(\ref{eq:conv_logi}), i.e.,
\begin{equation}
\frac{B^{\rm L}(\infty)}{B^{\rm L}(0)} = 1 + \frac{S^{\rm L}(0)}{B^{\rm L}(0)} \ ,
\end{equation}
where we have used the limit
\begin{equation}
\lim_{m\rightarrow0}\frac{\Gamma\left[m,2\ln(2^{m}\xi)\right]}{(m-1)!} = 0 \ .
\end{equation}  

\begin{figure}[h]
  \begin{center}
      \includegraphics[scale=0.9]{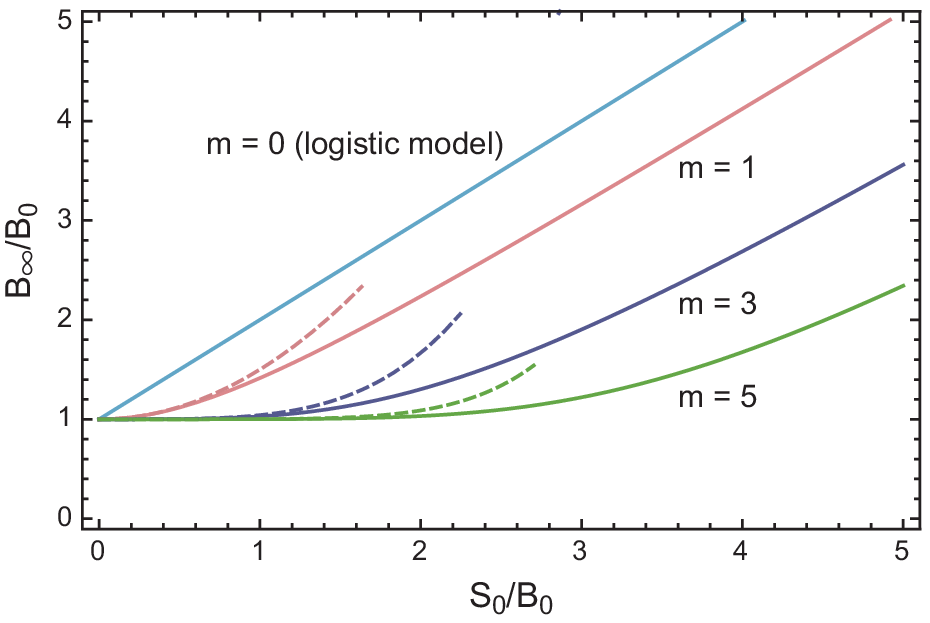}
\mbox{}
      \includegraphics[scale=0.9]{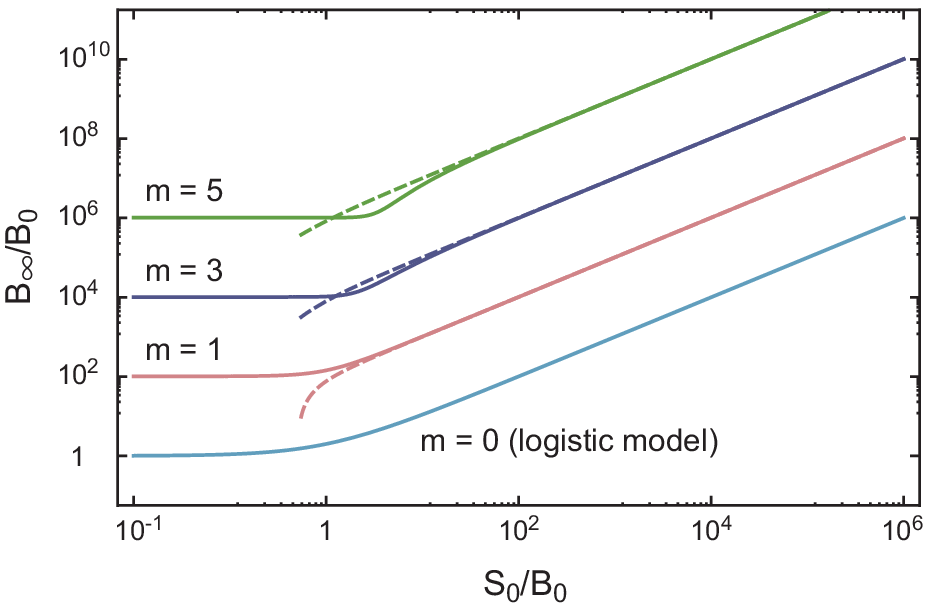}
    \caption{\label{fig:conv_asymp}(color online) The convergence value of the number of the bacteria over the initial value as a function with respect to $S_0/B_0$ for several values of $m$. The dashed curves represent the asymptotic expressions, i.e., Eqs.~(\ref{eq:Binf_small})  and (\ref{eq:Binf_large}). For clarity, each successive curve on the right plot is shifted two orders up.
}
  \end{center}
\end{figure}

Figure~\ref{fig:conv_asymp} shows $B_\infty/B_0$ as a function with respect to $S_0/B_0$ calculated numerically from Eq.~(\ref{eq:conv})  for several values of $m$. The asymptotic representations of Eqs.~(\ref{eq:Binf_small}) and (\ref{eq:Binf_large}) are also plotted as dashed curves. For clarity, on the right plot, each successive curve is shifted two orders up. We can see that the above discussions are correct from the degree to which the numeric solutions of the convergence values of NLM match the asymptotic representations.

\section{Fitting real data of bacterial growth}
\label{sec:fitting}

In this section, we shall fit real experimental data of bacteria growth using NLM, MGM, MLM, and the new logistic model (NewLM) \cite{Fujikawa03}, and compare the degree to which each model fits the data. 

The data are for growth curves of {\it Escherichia coli} and {\it Salmonella spp}, with the number of bacteria given per unit volume. Therefore, we utilize NLM as expressed by Eqs.~(\ref{eq:neologi3}). In addition, we use the following expressions for  MLM (Eq.~(\ref{eq:MLM2})) and MGM (Eq.~(\ref{eq:MGM2})).
\begin{equation}
{b^{\rm mL}(t)} = {b^{\rm mL}(0)}\exp\left[-\frac{\beta}{1+\delta}+\frac{\beta}{1+\delta{\rm e}^{-\gamma t}}\right] \ ,
\label{eq:MLM2}
\end{equation}
\begin{equation}
{b^{\rm mG}(t)} = {b^{\rm mG}(0)}\exp\left[-C\exp\left[-{\rm e}^{DM}\right]+C\exp\left[-\exp\left\{-D(t-M)\right\}\right]\right] \ ,
\label{eq:MGM2}
\end{equation}
where $b^{\rm mL}(t)\equiv B^{\rm mL}(t)/L^3$ and ${b^{\rm mG}(t)}\equiv {B^{\rm mG}(t)}/L^3$.

For the same reason, we use the following expression for NewLM.
\begin{equation}
\frac{{\rm d}b^{\rm nL}(t)}{{\rm d}t} = rb^{\rm nL}(t)\left\{1-\frac{b^{\rm nL}(t)}{b^{\rm nL}_\infty}\right\}
\left\{1-\frac{b^{\rm nL}_{\rm min}}{b^{\rm nL}(t)}\right\}^c \ ,
\label{eq:NewLM}
\end{equation}
where $b^{\rm nL}(t)$ represents the number of bacteria per unit volume at time $t$.
The other quantities are fitting parameters. When  $b^{\rm nL}(t)$ is close to $b^{\rm nL}_\infty$, the R.H.S of Eq.~(\ref{eq:NewLM}) converges to 0. Thus, $b^{\rm nL}_\infty$ represents the saturated number of bacteria per unit volume in the stationary phase. In Ref.~\cite{Fujikawa03}, the authors imposed the constraint between $b^{\rm nL}_{\rm min}$ and $b^{\rm nL}(0)$ as $b^{\rm nL}_{\rm min}=(1-10^{-6})b^{\rm nL}(0)$ when fitting the data for bacterial growth.

When finding the best fit parameters for a model, we minimize the residual sum of squares (RSS) between the logarithmic number of the data and the model.

\subsection{\it Escherichia coli}
\label{subsec:Ecoli}

Now, we shall analyze the fits of these models to the growth data of {\it E. coli} K12 MG1655 strains with 0.2 \% glucose and casamino acids from a previous study \cite{Madar13}. This data was kindly provided by A. Bren who is one of the authors this study. In order to measure the mean cell number of {\it E. coli}, they used flow cytometry \cite{Madar13}. 

The best fit parameters minimizing the RSS for NLM (for several values of $m$), NewLM, MGM, and MLM are shown in Table~\ref{Tab:Ecol_30}. From the RSSs in this table, we can see that NLM with $m = 3$ describes the data more precisely than MLM and MGM, and with slightly better precision than NewLM.

\begin{figure}[h]
  \begin{center}
      \includegraphics[scale=1.3]{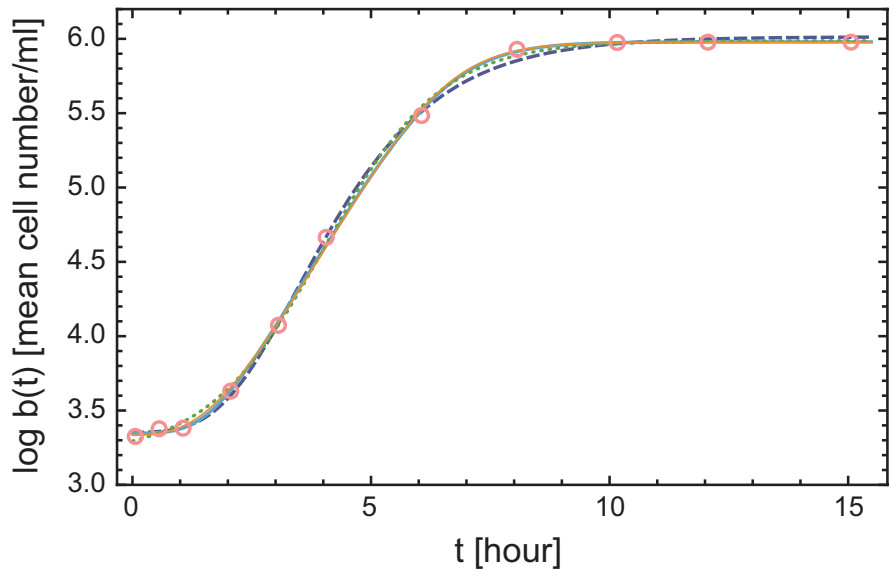}
    \caption{\label{fig:Ecoli_37}(color online) The temporal evolution of the logarithmic number of {\it E. coli} per unit ml with glucose (open circles) is shown with the best fits of this data for NLM ($m=3$, thick blue curve), NewLM (thin orange curve), MGM (dashed navy curve), and MLM (dotted green curve).}
  \end{center}
\end{figure}

In Figure~\ref{fig:Ecoli_37} the {\it E. coli } growth data (open circles) are shown with the best fits for this data of NLM (with $m = 3$, thick blue curve), NewLM (thin orange curve), MGM (dashed navy curve), and MLM (green dotted curve). Since NLM and NewLM curves coincide almost perfectly with each other, it may be difficult to distinguish them in this figure. NLM and NewLM are particularly more precise in describing the data in the transition region after the exponential phase, and before the stationary phase.
\begin{table}
\caption{\label{Tab:Ecol_30}Parameters for minimizing RSS of NLM, NewLM, MGM, and MLM for {\it E. coli} growth data with glucose.}
\begin{center}
\begin{tabular*}{1.0\textwidth}{p{0.15\textwidth}p{0.070\textwidth}p{0.070\textwidth}p{0.070\textwidth}p{0.070\textwidth}p{0.070\textwidth}p{0.070\textwidth}p{0.070\textwidth}p{0.070\textwidth}p{0.069\textwidth}p{0.069\textwidth}p{0.069\textwidth}}
\hline
                                                  &\multicolumn{8}{c}{NLM}   &                                                     \\
                                                  &$m=0$&$m=1$ &$m=2$    &$m=3$  &$m=4$ &$m=5$ &$m=6$    &$m=7$              & NewLM & MGM & MLM   \\
                                                  & (LM)   &            &               &             &           &             &               &                         &         &          &   \\
\hline
\hline
$\alpha\times10^{6}$ \ [ml/h]&$0.859$&$0.959$&$1.11$&$1.30$ & $1.50$&$1.73$&$1.98$ &$2.27$                     & --            & --        & -- \\
$b_0\times10^{-3}$\ [1/ml] &$1.14$  &$1.63$ & $1.98$ &$2.23$ & $2.41$ &$2.55$&$2.66$&$2.73$                      &$2.16$     &$2.25$ & $1.97$\\
$s_0\times10^{-6}$  \ [1/ml]&$1.06$ &$1.03$& $0.992$  &$0.958$ & $0.928$ &$0.899$&$0.841$&$4.05$            & --            & --         & -- \\
\hline
RSS                                         &$0.205$&$0.0915$& $0.0248$& $0.00607$&$0.0215$&$0.0572$&$0.102$&$0.149$&$0.00852$&$0.0102$& $0.0121$\\
\hline
\hline
\end{tabular*}
\end{center}
\end{table}

From Table~\ref{Tab:Ecol_30}, we see that the order of $\alpha$ is $10^{-6} \ \mbox{[ml/h]}$. 
We can estimate the average number of substrates required for cell divisions from this order in the following way. The volume of {\it E. coli} is approximately $10^{-18} \ [\mbox{m}^3]=10^{-12} \  [\mbox{ml}]$ \cite{Phillips12}. 
The swimming velocity of {\it E. coli} K12 MG1655 with glucose is approximated by using the swimming velocity of a closely related strain, {\it E. coli} K12 MTCC1302 with 2-Deoxy-D-glucose, which is $2\times10^{-5}\ [\mbox{m/s}] \approx 10^{-1} \ [\mbox{m/h}]$ \cite{Bhaskar15}. The length of a substrate cube required for cell division can be calculated by using the interpretation of $\alpha$, Eq.~(\ref{eq:interpretation_of_alpha}) as 
\begin{equation}
\mbox{(length of  a substrate cube)} \sim 10^{-7} \ [\mbox{m}]  \ .
\label{eq:interpretation_of_alpha3}
\end{equation}
The culture medium of the {\it E. coli} growth data used contains glucose as a substrate. Here we roughly estimate the volume of a glucose molecule as $(10^{-10})^3 \ [\mbox{m}^3] = 10^{-30} \ [\mbox{m}^3]$. Therefore, the number of glucose molecules required for cell division is roughly $10^9$ because the volume of space the glucose molecules occupy is estimated to be $10^{-21}\ [\mbox{m}^3]$ from Eq.~(\ref{eq:interpretation_of_alpha3}). 

We can obtain the number of glucose molecules required for cell divisions from {\it E. coli} yield on glucose whose average value is 0.5 g/1 g glucose \cite{Shiloach05}. From Fig.~\ref{fig:Ecoli_37}, the generation time can be estimated as approximately $30 \ [\mbox{min}]$ because the value of $\log b(t)$ increases by approximately 0.6 (nearly equal to $\log 4$) in the time interval from 3 to 4 hours after beginning the culture. 
Therefore, the dry mass of one {\it E. coli }bacterium is roughly $6\times10^{-13} \ [\mbox{g}]$ \cite{Neidhardt96}, and so approximately $10^{-12} \ [\mbox{g}]$ of glucose are required for one {\it E. coli} bacterium cell division. From this, we can estimate that $10^9$ glucose molecules are required for a single {\it E. coli} bacterium to divide by using the molecular weight of glucose,  $180 \ [\mbox{g/mol}]$. 
Thus, the number of glucose molecules required for cell division, as estimated by NLM, does not differ from that calculated by empirical methods, even though NLM utilizes a simplified idealization of substrates.

\subsection{\it Salmonella spp.}

Here, we present the results of fitting real data for the growth of {\it Salmonella spp.}, cultured by the Institute of Food Research, UK (obtained through ComBase, https://www.combase.cc/), with NLM, NewLM, MGM, and MLM. The culture medium contained tryptone soya broth (TSB), and the number of {\it Salmonella spp.} cells was obtained by colony-forming-unit (CFU) measurements.

Table~\ref{Tab:Sal_30_6.7_0.974} shows the fit parameters minimizing the RSS for NLM (for several values of $m$), NewLM, MGM, and MLM for the growth data of {\it Salmonella spp.} in TSB at $30{}^\circ$C, pH 6.7 and aw 0.974. From the RSSs in the table, we find that NLM (with $m = 5$) provides the best fit for the real data, yielding a significantly more precise fit than MGM and MLM, and a slightly more precise fit than NewLM.

\begin{figure}[h]
  \begin{center}
      \includegraphics[scale=1.3]{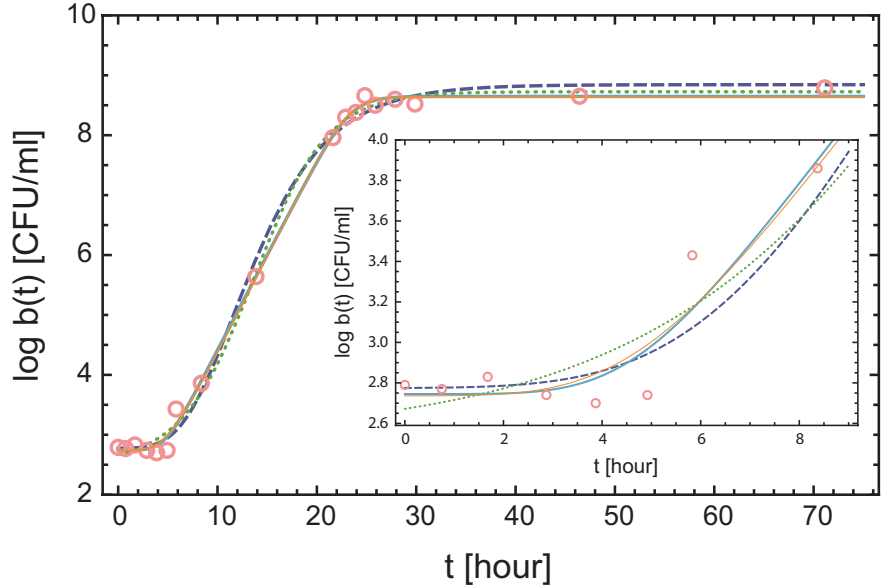}
    \caption{\label{fig:Sal_30_6.7_0.974}(color online) The temporal evolution of the logarithmic number of {\it Salmonella spp.} per unit ml in TSB at $30{}^\circ$C, pH 6.7, and aw 0.974 (open circles) is shown with the best fits of this data for NLM ($m=5$, thick blue curve), NewLM (thin orange curve), MGM (dashed navy curve), and MLM (dotted green curve).}
  \end{center}
\end{figure}

This temporal evolution of the logarithmic number of {\it Salmonella spp.} per unit ml is shown in Fig.~\ref{fig:Sal_30_6.7_0.974} by open circles. The best fits for NLM, NewLM, MGM and MLM are represented in this figure by the thick, thin, dashed and dotted curves, respectively. Again, the curves of NLM and NewLM are so close to each other that it is difficult to distinguish them, reflecting their nearly identical values of RSS. In the inset of Fig.~\ref{fig:Sal_30_6.7_0.974}, which shows the bacterial growth around the lag phase, we can see the minor differences between them, and MLM cannot describe the lag phase precisely.
\begin{table}
\caption{\label{Tab:Sal_30_6.7_0.974}Parameters for minimizing RSS of NLM, NewLM, MGM, and MLM for {\it Salmonella spp.} growth data in TSB at $30{}^\circ$C, pH 6.7, and aw 0.974.}
\begin{center}
\begin{tabular*}{1.0\textwidth}{p{0.15\textwidth}p{0.070\textwidth}p{0.070\textwidth}p{0.070\textwidth}p{0.070\textwidth}p{0.070\textwidth}p{0.070\textwidth}p{0.070\textwidth}p{0.070\textwidth}p{0.069\textwidth}p{0.069\textwidth}p{0.069\textwidth}}
\hline
                               &\multicolumn{8}{c}{NLM}   &                                                     \\
                               &$m=0$&$m=1$ &$m=2$    &$m=3$  &$m=4$ &$m=5$ &$m=6$    &$m=7$                & NewLM & MGM & MLM   \\
                               & (LM)   &            &               &             &           &             &               &                           &             &          &   \\
\hline
\hline
$\alpha\times10^{9}$ \ [ml/h]&$1.17$&$1.22$& $1.28$&$1.35$ & $1.50$&$1.65$   &$1.81$   &$1.99$ & --         & -- & -- \\
$b_0$                       \ [1/ml] &$121$  &$198$  & $290$    & $387$ & $476$ &$556$  &$625$   &$683$  & $545$  &$596$ & $470$\\
$s_0\times10^{-8}$  \ [1/ml]&$5.23$&$5.15$  & $5.02$ &$4.96$ & $4.66$ &$4.44$  &$4.24$  &$4.05$  & --       & --        & --\\
\hline
RSS                       &$1.91$    &$1.39$& $0.876$& $0.493$&$0.278$&$0.216$&$0.263$&$0.378$              &$0.230$&$0.474$ & $0.362$\\
\hline
\hline
\end{tabular*}
\end{center}
\end{table}

The best fit parameters minimizing the RSS for NLM (for several values of $m$), NewLM, MGM, and MLM for {\it Salmonella spp.} growth data in TSB at $30{}^\circ$C, pH 5.3 and aw 0.997 are listed in Tab.~\ref{Tab:Sal_30_5.3_0.997}. 
Here, NLM with $m = 5$ also provides a much better fit than MLM and MGM. However, NewLM provides a slightly better fit than NLM in this case.

\begin{figure}[h]
  \begin{center}
      \includegraphics[scale=1.3]{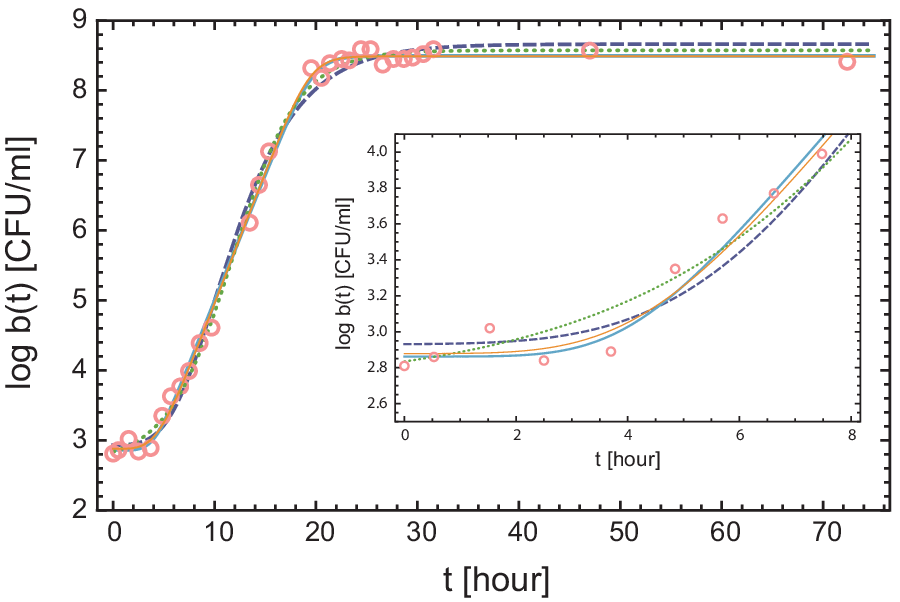}
    \caption{\label{fig:Sal_30_5.3_0.997}(color online) The temporal evolution of the logarithmic number of {\it Salmonella spp.} per unit ml in TSB at  $30{}^\circ$C, pH 5.3, and aw 0.997 (open circles) is shown with the best fits of this data for NLM ($m=5$, thick blue curve), NewLM (thin orange curve), MGM (dashed navy curve), and MLM (dotted green curve).}
  \end{center}
\end{figure}

The temporal change of the logarithmic number of {\it Salmonella spp.} per unit ml is shown in Fig.~\ref{fig:Sal_30_5.3_0.997} by open circles. The best fits for NLM, NewLM, MGM and MLM are represented in this figure by the thick, thin, dashed and dotted curves, respectively. Again, the fits for NLM and NewLM are too close to each other to be easily distinguished.
\begin{table}
\caption{\label{Tab:Sal_30_5.3_0.997}Parameters for minimizing RSS of NLM, NewLM, MGM, and MLM for {\it Salmonella spp.} growth data in TSB at $30{}^\circ$C, pH 5.3 and aw 0.997.}
\begin{center}
\begin{tabular*}{1.0\textwidth}{p{0.15\textwidth}p{0.070\textwidth}p{0.07\textwidth}p{0.07\textwidth}p{0.07\textwidth}p{0.07\textwidth}p{0.07\textwidth}p{0.07\textwidth}p{0.07\textwidth}p{0.069\textwidth}p{0.069\textwidth}p{0.069\textwidth}}
\hline
                                  &\multicolumn{8}{c}{NLM}   &                                                     \\
                                  &$m=0$&$m=1$ &$m=2$  &$m=3$   &$m=4$ &$m=5$ &$m=6$    &$m=7$                    & NewLM & MGM & MLM   \\
                                  & (LM)   &            &            &               &           &             &               &                                &         &          &   \\
\hline
\hline
$\alpha\times10^{9}$ \ [ml/h]&$2.02$&$2.09$ & $2.19$  &$2.33$& $2.49$&$2.68$   &$2.88$   &$3.09$       & --  & -- & -- \\
$b_0$                       \ [1/ml]  &$143$  &$235$  & $348$  & $473$   & $602$ &$728$   &$853$   &$972$        & $757$&$853$ & $683$\\
$s_0\times10^{-8}$ \ [1/ml]  &$3.42$&$3.38$  & $3.33$ &$3.26$ & $3.19$ &$3.12$  &$3.05$  &$3.00$        & --       & --        & --\\
\hline
RSS                             &$2.24$ &$1.70$& $1.16$& $0.723$&$0.452$&$0.341$&$0.359$&$0.470$                    &$0.313$&$0.793$ & $0.420$\\
\hline
\hline
\end{tabular*}
\end{center}
\end{table}

From Tables~\ref{Tab:Sal_30_6.7_0.974} and \ref{Tab:Sal_30_5.3_0.997}, $\alpha$ is roughly of the order $10^{-9} \ [\mbox{ml/h}]$. Unlike Subsection~\ref{subsec:Ecoli}, we cannot estimate the number of substrate molecules required for {\it Salmonella spp.} division from this order of $\alpha$.
That is because a measurement of the growth yield for {\it Salmonella spp.} on TSB is not possible as TSB is a mixture of protein degradation products, and its molecular components are unclear. 
However, according to Eq.~(\ref{eq:interpretation_of_alpha}), we can interpret the lower order of $\alpha$ for {\it Salmonella spp.} than {\it E. coli} is simply the result of the fact that the volume of protein degradation product required for {\it Salmonella spp.} cell division is larger than the volume of glucose required for cell division of {\it E. coli.}

\begin{figure}[h]
  \begin{center}
      \includegraphics[scale=1.2]{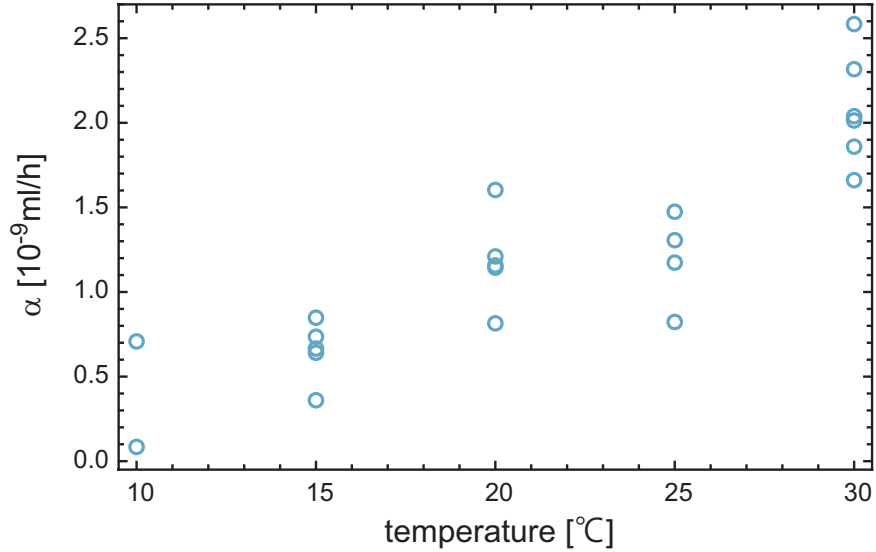}
    \caption{\label{fig:alpha_dependence}(color online) The temperature dependence of $\alpha$  derived by fitting {\it Salmonella spp.} growth data where pH is in the range from 5.5 to 6.5.}
  \end{center}
\end{figure}

Finally, we shall discuss the dependence of $\alpha$ on temperature. Figure~\ref{fig:alpha_dependence} shows this dependency derived by fitting 22 different {\it Salmonella spp.} data sets where pH ranges from 5.5 to 6.5. The temperature dependence of the number of protein degradation product molecules required for cell division, and the temperature dependence of the volume of {\it Salmonella spp.} cells are negligibly small compared with the temperature dependence of the velocity. Therefore, this implies that the moderately positive temperature dependence of $\alpha$ originates from the temperature dependence of velocity. This result is not contradictory to experiments with other bacteria showing such a positive dependency \cite{Lowe87,Miyata02,Oleksiuk11,Humphries13}.

\section{Prediction of the saturated number of bacteria in stationary phase from initial data}
\label{sec:prediction}

Now, we will predict the total number of bacteria in the stationary phase by using the initial data including, and just after the end of, the lag phase, whose time interval is $0\le t\le t_{\rm c}$. The real data of bacterial growth that we shall use are the same as that used in Sec.~\ref{sec:fitting}. 

We obtain the saturated number of bacteria (SNB) per unit volume in the stationary phase for NLM by dividing Eq.~(\ref{eq:conv}) by $L^3$ as follows:
 \begin{equation}
b_\infty \equiv B_\infty/L^3 = F_m\left[G_m^{-1}\left(s_0/b_0\right)\right]s_0 \ .
\label{eq:conv2}
\end{equation}
The SNB per unit volume for MLM and MGM can be derived by taking the limit $t\rightarrow\infty$ in Eqs.~(\ref{eq:MLM2}) and (\ref{eq:MGM2}) as
\begin{equation}
{b^{\rm mL}_\infty} \equiv \lim_{t\rightarrow\infty}{b^{\rm mL}(t)} = {b^{\rm mL}(0)}{\rm e}^\beta \ , \ \
{b^{\rm mG}_\infty} \equiv \lim_{t\rightarrow\infty}{b^{\rm mG}(t)} = {b^{\rm mG}(0)}{\rm e}^C \ .
\end{equation}
NewLM includes SNB explicitly as one of the fitting parameters.

The accuracy of the prediction of SNB is evaluated by comparing the relative difference between the logarithmic SNB derived from the initial data, and the logarithmic SNB derived from the full data set. The SNBs are calculated from the best fit parameters that minimize RSS. Note that there are several local minima for RSS in all the models. In the case of NLM, we adopt the best fitting parameter set for which the order of $\alpha$  is within approximately one order of magnitude of that obtained in  Sec.~\ref{sec:fitting}, because the value of  $\alpha$ defined in Eq.~(\ref{eq:interpretation_of_alpha}) should be determined by the types of bacteria and substrates, and a large deviation in the value from that obtained in Sec.~\ref{sec:fitting} is unphysical.

\begin{figure}[h]
  \begin{center}
      \includegraphics[scale=0.9]{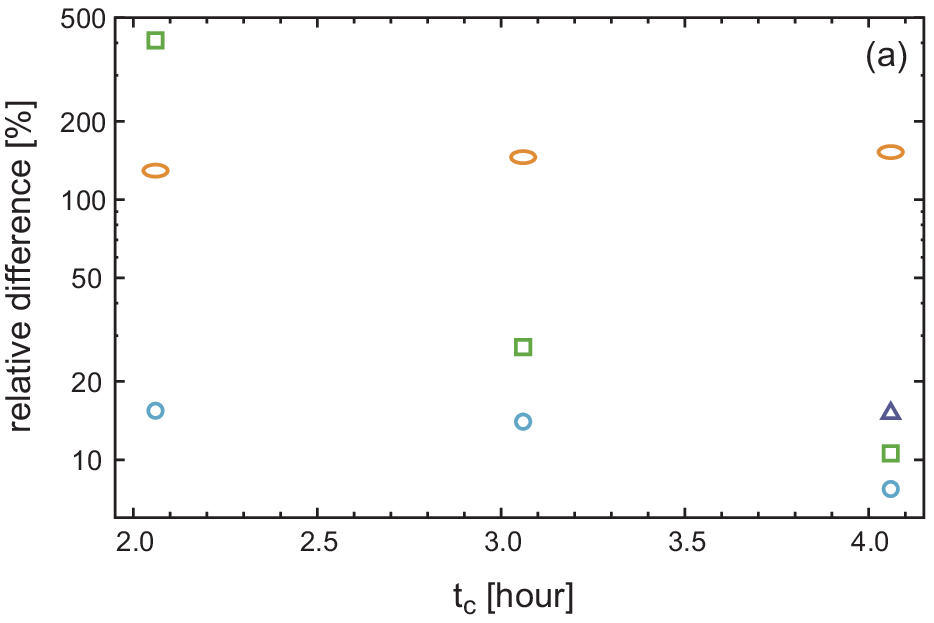}
      \includegraphics[scale=0.95]{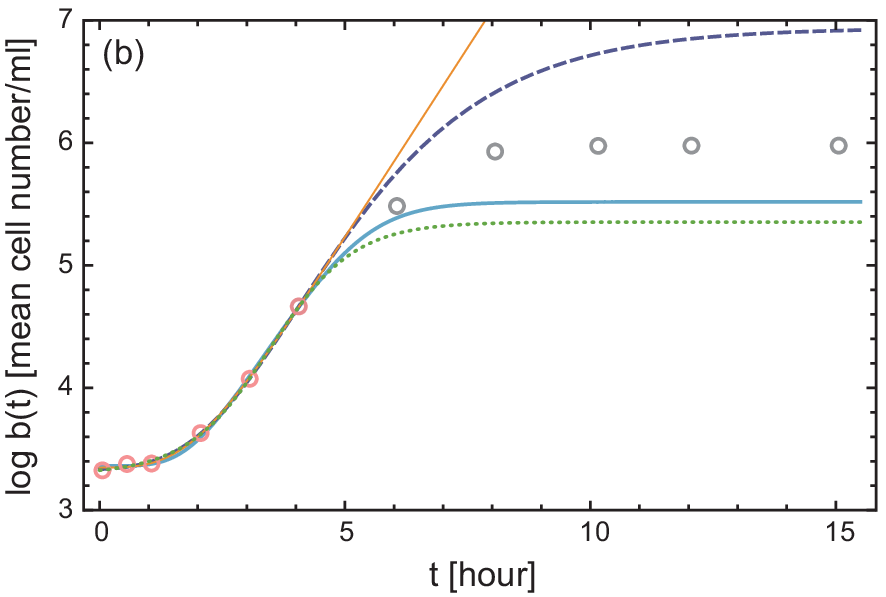}
    \caption{\label{fig:relative_error:E.coli}(color online) (a) The relative differences of the logarithmic SNB of {\it E. coli} predicted from the initial data, and from the full data set, for NLM (blue circle), NewLM (orange ellipse), MGM (navy triangle), and MLM (green square). (b) Best fits of NLM ($m=4$, thick blue curve), NewLM (thin orange curve), MGM (dashed navy curve), and MLM (dotted green curve) to the initial 6 data points of the E. coli growth data (circles) for (a).}
  \end{center}
\end{figure}

The relative differences of the logarithmic SNB of {\it E. coli} predicted from initial data, and from the full data set, for NLB, NewLM, MGM, and MLM for various time intervals, $t_c$, are shown in Fig.~\ref{fig:relative_error:E.coli}(a) by circles, ellipses, triangles, and squares, respectively. We do not show the relative differences of MGM for $t_c=2.06$ and $3.06$ because the values are respectively $2.67\times10^5$ \% and $1.18\times10^{-3}$ \%, respectively, and plotting these differences would obscure the relative differences of the other models.

From this figure, we can see that NLM has relative differences below approximately 15 \%, in contrast to the relative differences for the other models which can exceed 100 \%.
Figure~\ref{fig:relative_error:E.coli}(b) shows the models' predictions of the logarithmic SNB from the initial 6 data points corresponding to $t_c=4.06$.

\begin{figure}[h]
  \begin{center}
      \includegraphics[scale=0.9]{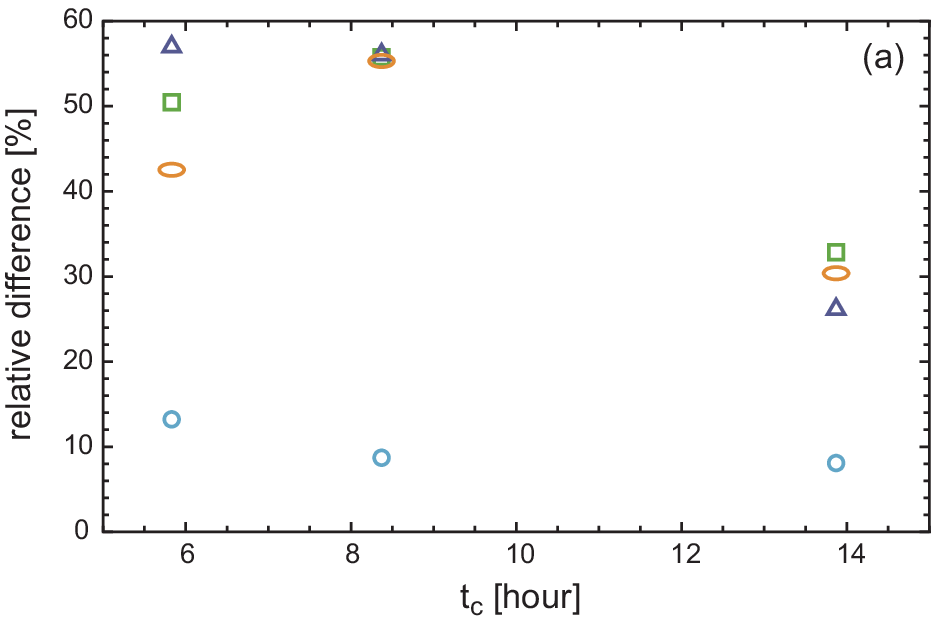}
      \includegraphics[scale=0.95]{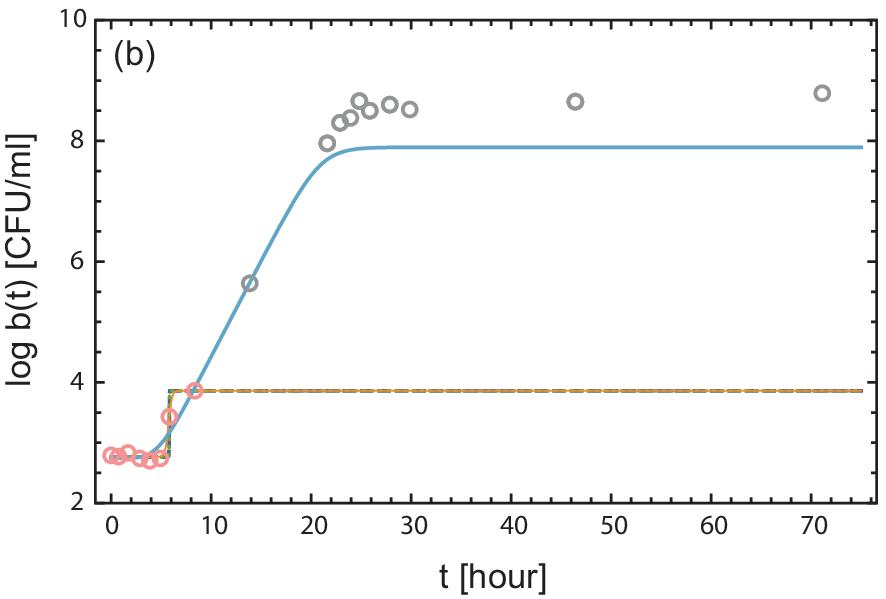}
    \caption{\label{fig:relative_error:S.spp:pH6.7}(color online) (a) The relative differences of the logarithmic SNB of {\it Salmonella spp.} in TSB at $30{}^\circ$C, pH 6.7 and aw 0.974 predicted from the initial data, and from the full data set, for NLM (blue circle), NewLM (orange ellipse), MGM (navy triangle), and MLM (green square). (b) Best fits of NLM ($m=5$, thick blue curve), NewLM (thin orange curve), MGM (dashed navy curve), and MLM (dotted green curve) to the initial 8 data points of the {\it Salmonella spp.} growth data (circles) for (a).}
  \end{center}
\end{figure}

\begin{figure}[h]
  \begin{center}
     \includegraphics[scale=0.9]{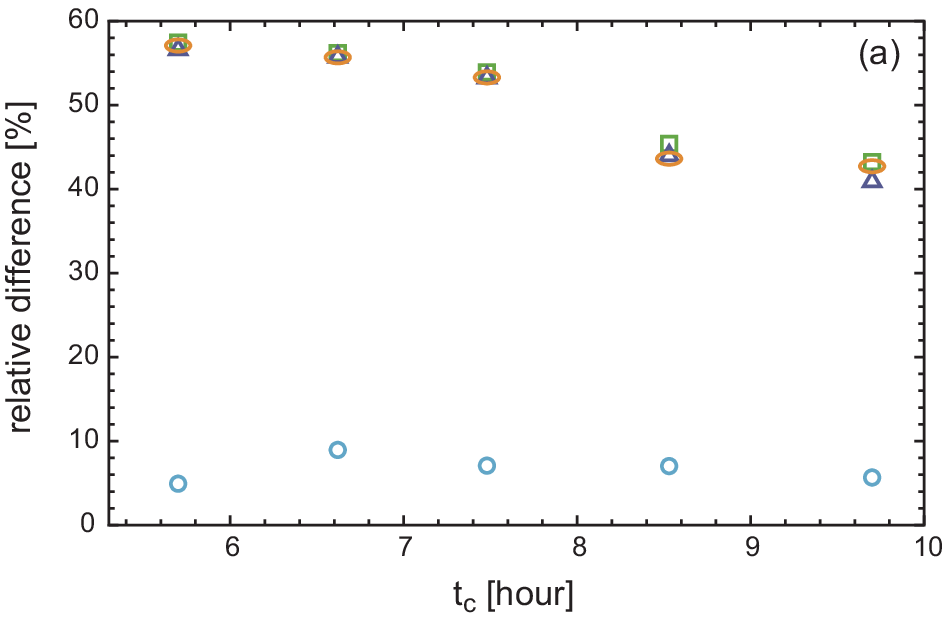}
     \includegraphics[scale=0.95]{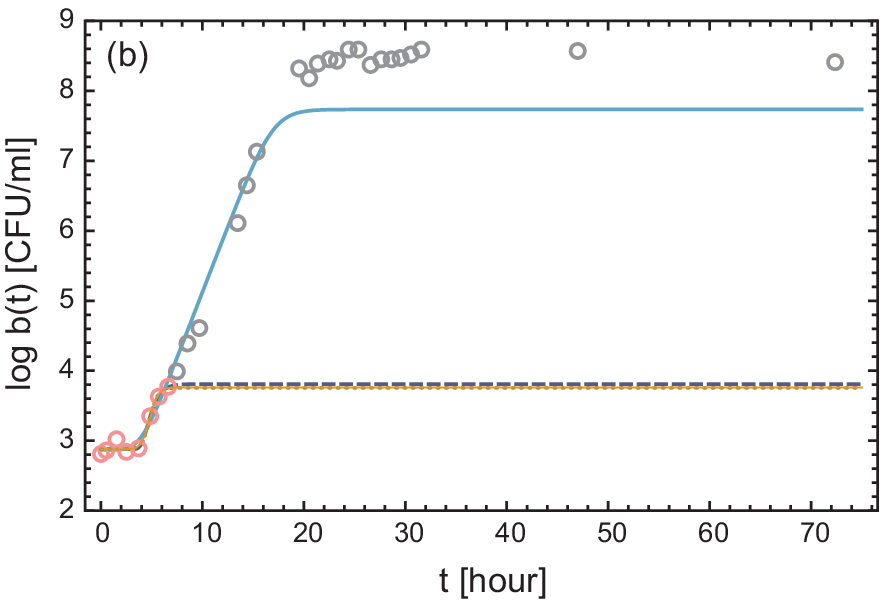}
    \caption{\label{fig:relative_error:S.spp:pH5.3}(color online) (a) The relative differences of the logarithmic SNB of {\it Salmonella spp.} in TSB at $30{}^\circ$C, pH 5.3 and aw 0.997 predicted from the initial data, and from the full data set, for NLM (blue circle), NewLM (orange ellipse), MGM (navy triangle), and MLM (green square). (b) Best fits of NLM ($m=5$, thick blue curve), NewLM (thin orange curve), MGM (dashed navy curve), and MLM (dotted green curve) to the initial 8 data points of the {\it Salmonella spp}. growth data (circles) for (a).}
  \end{center}
\end{figure}

Figures ~\ref{fig:relative_error:S.spp:pH6.7}(a) and \ref{fig:relative_error:S.spp:pH5.3}(a) show the relative differences of the logarithmic SNB of {\it Salmonella spp.} at $30{}^\circ$C, pH 6.7, and aw 0.974, and at $30{}^\circ$C, pH 5.3 and aw 0.997, respectively, predicted from initial data, and from the full data set, for NLB (circles), NewLM (ellipses), MGM (triangles), and MLM (squares), for various time intervals, $t_c$. The relative differences for NLM are less than 10 \%.
In contrast, the relative errors for NLM,  MGM, and MLM are approximately 50 \%. In Figs.~\ref{fig:relative_error:S.spp:pH6.7}(b) and \ref{fig:relative_error:S.spp:pH5.3}(b), we can clearly see the difference in accuracy in the prediction of the logarithmic SNB, using the initial 8 data points, between NLM and the other models.

From this analysis, the differences in the predictions of the logarithmic SNB from the initial data, and predictions from the entire data set are the least for NLM, with the exception of the data of  {\it E. coli} with $t_c=3.06$. Note that the relative differences are calculated for the logarithmic SNB. We cannot predict the SNB in stationary phase precisely using NLM, but can only estimate the order of the SNB. In contrast, the other models are useless for such an estimation, because of the large relative differences.

\section{Concluding remarks}
\label{conclude}

In this paper, we constructed NLM that describes bacterial growth curves in the logarithmic scale precisely. 

Either LM or GM cannot approximate the lag phase of growth data on the logarithmic scale. There are two approaches that allow the LM and GM to approximate the lag phase. The first is to take the exponential of each model. The models yielded in this way are MLM and MGM. The second is to multiply the right hand side of Eq.~(\ref{eq:LM}) by an extra term to obtain the NewLM, Eq.~(\ref{eq:NewLM}).

Our approach is distinguished from those of the models, and is rooted in actual physical phenomena. We derived NLM in the following way. First, we revealed elementary processes that could be used to derive the LM. However, the synthesis of inducible enzymes was not considered in these elementary processes. Thus, in order to more accurately describe bacterial growth, we constructed NLM by incorporating the mechanisms related to the production of inducible enzymes into LM.

We then fitted {\it E. coli} and {\it Salmonella spp.} growth data by NLM, NewLM, MGM, and MLM, and compared the degree to which these models fit the data. From this analysis, it is clear that NLM approximates real growth data slightly better than NewLM, and much better than MLM and MGM. Furthermore, we estimated the number of glucose molecules required for {\it E. coli} cell division using a value of  $\alpha$ derived by fitting {\it E. coli} data with NLM, and using the definition of  $\alpha$, Eq. (24), and this estimation matched that obtained by experimental data. 
This confirms the validity of our model, and makes the meaning of  $\alpha$ physically clear. By contrast, other models do not have such a physically relevant parameter, because they are derived by modifying LM or GM formally.

Equation~(\ref{eq:interpretation_of_alpha})  implies that the growth rate for bacteria becomes lower as the volume of substrates required for cell division increases for the same bacteria and environment. This can be interpreted to mean that as the volume of substrates required for cell division increases, so does the time required for their assimilation. An experiment supporting this concept has been reported \cite{Bren16}. {\it E. coli}  strains in a culture of NH$_4$Cl and maltotriose (consisting of three glucose molecules) grows slower than in a culture with glucose. 
However, there are exceptions. When a single amino acid, e.g. arginine, glutamate, or proline, is the sole nitrogen source, maltotriose provides faster growth than glucose \cite{Bren16}. This counterintuitive growth is due to the build-up of TCA cycle intermediates, which lead to the inhibition of cAMP in an environment with glucose and poor nitrogen sources \cite{Bren16}. Since NLM does not include the TCA cycle and variables related cAMP, it cannot describe such a rare situation.

We also attempted to predict the logarithmic SNB in the stationary phase from the initial data by fitting the data using NLM and the other models. We showed that the relative differences of the logarithmic SNB predicted from initial data, and the logarithmic SNB predicted from the entire data set, for NLM are less than 15 \%, while the relative differences in the other models are much higher. Therefore, we can use NLM for estimating the order of SNB in the stationary phase from initial data.

The good estimation of the order of SNB by NLM can be understood by the physical relevance of $\alpha$. 
The value of $\alpha$ as defined by Eq.~(\ref{eq:interpretation_of_alpha}) is determined by the types of bacteria and substrates. Thus, we can narrow the range of $\alpha$ while predicting SNB from the initial data. Another reason might be the formulation of SNB,
\begin{equation}
{b_\infty} \simeq \left\{1-\frac{\Gamma\left[m,2\ln(2^{m}s_0/b_0)\right]}{(m-1)!}\right\}{b_0}
\end{equation}
for $s_0/b_0\gg1$.
This is formulated from the initial numbers of bacteria and substrates, and $m$, which is related to the length of the lag phase. In this way, NLM can be used to predict SNB from the initial data including, and just after the end of, the lag phase. 
On the other hand, the other models use the SNB as one of their fitting parameters.

\subsection{Acknowledgments}
We would like to thank K. Yura, S. Tsuru, E. Foley, and the members of astrophysics laboratory at Ochanomizu
University for extensive discussions.


\end{document}